\documentclass[a4paper, amsfonts, amssymb, amsmath, reprint, showkeys, nofootinbib, twoside, aps, pra]{revtex4-2}
\usepackage[english]{babel}
\usepackage[utf8]{inputenc}
\usepackage[colorinlistoftodos, color=green!40, prependcaption]{todonotes}
\usepackage{amsthm}
\usepackage{mathtools}
\usepackage{physics}
\usepackage{xcolor}
\usepackage{graphicx}
\usepackage[left=23mm,right=13mm,top=35mm,columnsep=15pt]{geometry}
\usepackage{adjustbox}
\usepackage{placeins}
\usepackage[T1]{fontenc}
\usepackage{lipsum}
\usepackage{csquotes}
\usepackage{subcaption}
\usepackage{float}
\usepackage{amsmath}
\usepackage{ulem}
\newcommand{\dgr}{}
\newcommand{\lgr}{}
\usepackage[pdftex, pdftitle={Article}, pdfauthor={Author}]{hyperref} 
\begin{document}
\title{Modelling of transient interference phenomena in collinear laser spectroscopy}
\author{Jovan Jovanovi\'c}
    \email[Email: ]{jovan.jovanovic@physics.ox.ac.uk}
    \affiliation{Theoretical Physics, Oxford University, 1 Keble Road, Oxford OX1 3NP, United Kingdom}

\author{Ronald Fernando Garcia Ruiz}
    \email[Email: ]{rgarciar@mit.edu}
    \affiliation{Massachusetts Institute of Technology,~Cambridge,
~MA~02139,~USA}

\date{\today} 

\begin{abstract}
Collinear laser spectroscopy of fast atomic beams 
has been established as one of the main tools to perform precision experiments with atoms containing short-lived nuclei. Although highly sensitive, the spectral resolution of these techniques is typically limited to several MHz. Here, we study the use of transient interference phenomena to potentially improve the experimental resolution.  
Previous attempts to implement Ramsey-type measurements with fast beams were limited by the lack of understanding of the observed line shapes. By using a simple model, we provide a satisfactory description of the previously observed line shapes, and propose alternative experimental schemes to improve the resolution in fast ion-beam experiments.
  
\end{abstract}


\maketitle

\section{Introduction} \label{sec:introduction}

Collinear laser spectroscopy techniques are considered as powerful tools to perform systematic studies of atoms and molecules  containing isotopes with short-lived unstable nuclei \cite{camp16,geo19,ver20,gar20,kau20,udr21,Yang22}. As the Doppler width, as seen along the axis of the motion of the atoms, is inversely proportional to the kinetic energy of the atoms \cite{camp16}, narrow Doppler width  measurements of atomic spectra can be achieved by electrostatic acceleration. 
Thermal Doppler widths of the order of GHz exhibited while at rest can be reduced to a few tens of MHz by accelerating the ions to a few tens of keV.
This resolution allows for precise isotope shift measurements of atomic spectra, where changes in the root-mean-square nuclear charge radii can be obtained along isotopic chains. 
Moreover, the nuclear spins, magnetic dipole, and electric quadrupole moments can be extracted from hyper fine structure measurements. 
These techniques can now be applied to study isotopes produced in quantities lower than just a few hundred ions per second \cite{gar16,gro20,koz21}, and with lifetimes as short as a few milliseconds \cite{far16,gro20}.

In parallel to the experimental progress, the rapid increase of computer power and the development of quantum many-body methods have enabled computations of nuclear and atomic properties with unprecedented precision \cite{Her20,Sahoo_2020,Vernon2022}.
 The theoretical accuracy achieved in the last few years is now reaching the precision of collinear laser spectroscopy experiments \cite{Maa19,Hey21,koz21,Oha22}. Moreover, accurate calculations of higher order nuclear properties can now be achieved \cite{rei20,gro22}.  
 Hence improving the experimental precision, while preserving high-efficiency, is becoming a major motivation of modern experimental developments \cite{gro22}.

Conventional collinear laser spectroscopy techniques allows measurements of atomic spectra with typical resolutions on the order of a few tens of MHz \cite{gro15,phi19}. 
 Improving the resolution and precision of these techniques can provide access to higher-order terms of the nuclear distribution, e.g. the fourth power of the nuclear charge radius \cite{rei20,flambaum2018isotope, papoulia2016effect}, and high-order moments of the nuclear charge distribution \cite{gro21}. 
Furthermore, precision isotope shift measurements have been proposed as a viable route to constrain the existence of possible new forces and particles\cite{berengut2018probing}. This has motivated numerous experimental and theoretical developments in high-precision isotope shift spectroscopy using cold, trapped atoms and ions \cite{flambaum2018isotope, breveman2019, frugiuele2017}. However, with a few exceptions \cite{wan04}, trapping techniques are not in general applicable to short-lived nuclei, which are produced in small quantities and with lifetimes that can be as short as a fraction of a second \cite{koz21,far16}. In this work we explore the use of transient interference phenomena to study fast ion beams. These techniques can be implemented to combine the high resolution of time-dependent methods with the high sensitivity offered by collinear laser spectroscopy.


\section{Transient interference in collinear laser spectroscopy} \label{sec:back}

\begin{figure}[]
  \includegraphics[width=\linewidth]{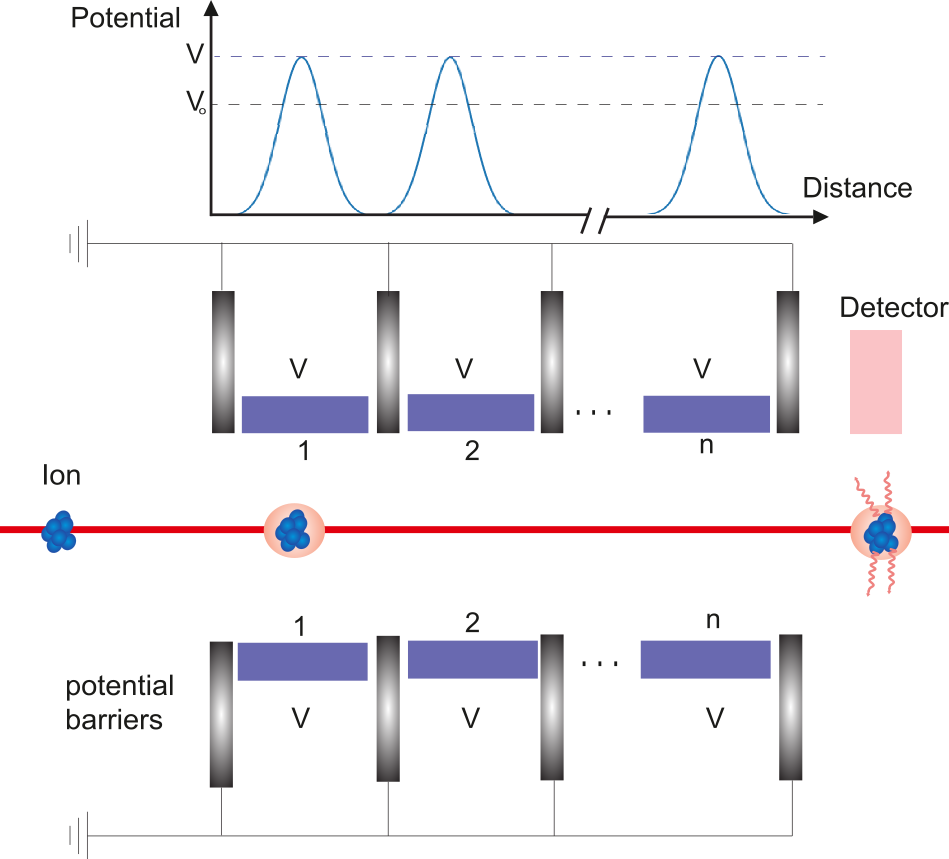}
  \caption{Scheme for Doppler switching of an ion beam passing through $n$ potential barriers.  The electrical potentials change the ion energy, thus, the frequency seen by the ions can be Doppler shifted around resonance (at potential $V_o$).}
  \label{fig:protocols}
\end{figure}

Transient interference phenomena in optical transitions have long been recognized as a potent method to achieve high-resolution measurements well below the natural linewidth of atomic transitions \cite{PhysRev.78.695,BORDE198910, Diddams825, PhysRevLett.104.070802}. 
In this work we focus on the study of a near-resonance laser field interaction with an ion travelling through different potential barriers. A scheme of the experimental sequence is shown in Figure \ref{fig:protocols}.
The laser field is detuned from the transition frequency in the ions' rest frame. The velocity of the ion is modified by passing through $n$ potential barriers, and tuned into resonance at a potential $V_0$. Hence, each potential barriers is analogous to an interaction region/period of a Ramsey-type measurement. When varying the distance between two potentials, the population of the excited state oscillates with a generalised Rabi frequency, from which the detuning, and hence the resonance frequency, can be extracted. 

This type of experiments were employed in collinear laser spectroscopy more than four decades ago, using one potential \cite{bor81}, and two potential barriers \cite{sil81}. These results highlighted that the inherent high time resolution achievable by using fast beams can be advantageous for  spectroscopy techniques \cite{bor81,bor84}. 
However, the Ramsey fringes that were experimentally observed \cite{sil81,bor81} exhibited relatively complex lineshapes that were not fully understood, limiting the applicability of the techniques. In the following section we use a simple model to study the interaction of a co-propagating laser and ion beam, passing through $n$ potential barriers. This simple model provides a satisfactory description of previous experiments using one potential \cite{bor81}, and two potential barriers \cite{sil81}.

\section{Two-level model} \label{sec:theory}
Considering a two-level atomic system coupled to a single electromagnetic (EM) mode, we can account for the dominant physical processes: stimulated emission, absorption and incoherent damping (e.g. spontaneous emission). A similar model was used to study optical nutation in collinear fast beam experiments \cite{wannstrom1990optical}. This system is described by a total Hamiltonian:
\begin{widetext}
\begin{equation}
\frac{1}{\hbar}\hat{H}(t) = \omega_{12} \ket{2}\bra{2} +  \omega\hat{a}^\dag \hat{a} + \frac{1}{\hbar}\sum_{\mu, \nu} \boldsymbol{F}\cdot \boldsymbol{D}_{\mu,\nu}i(\hat{a} e^{i\omega t} - \hat{a}^\dag e^{-i\omega t})\ket{\nu}\bra{\mu},
\end{equation}
\end{widetext}

~\\ where the two states are labelled as $\{\mu, \nu\}\in\{1,2\}$, $\hat{a}$ is the annihilation operator for the photons in the mode coupled to the system having frequency $\omega$. The energy difference between the two states is denoted by $\hbar\omega_{12}$. The last term is the interaction Hamiltonian.  $\boldsymbol{D}_{\nu,\mu}$ represent the electric dipole matrix elements in the basis of the two states, and $\boldsymbol{F} = \sqrt{\frac{\omega}{2\epsilon_0\hbar V}}\boldsymbol{\textbf{e}}$ characterises the electric field vector amplitude of the mode. 

This Hamiltonian acts on the states in the space:

\begin{equation}
    \ket{\Psi(t)} = \sum_{\mu}\ket{\xi_\mu(t)}\ket{\mu},
\end{equation}

~\\ where coefficients $\ket{\xi_\mu(t)}$ are vectors from the Hilbert space of the EM field. 
The equation of motion that governs the time evolution of the total density matrix of the system,
\begin{equation}
    i\hbar\frac{\partial}{\partial t} \ket{\Psi(t)}\bra{\Psi(t)} = i\hbar\frac{\partial}{\partial t} \hat{\rho}(t) = \comm{\hat{H}(t)}{\hat{\rho}(t)},\label{eq:Heis}
\end{equation}
can be expressed in a rotating frame via the transformation $\hat{U}_\omega(t) = (\ket{1}\bra{1} + e^{-i\omega t}\ket{2}\bra{2})\otimes e^{-it\omega\hat{a}^\dag\hat{a}}$:

\begin{equation}
\hat{H}_\omega(t) = \hat{U}_\omega(t)\hat{H}(t)\hat{U}^\dag_\omega(t) - i\hbar\frac{\partial\hat{U}_\omega(t)}{\partial t}\hat{U}^\dag_\omega(t).
\end{equation}

Working in this frame we can perform the usual steps in deriving the Lindblad-type Master equation \cite{2020} for the reduced density matrix of the two-level system:
\begin{equation}
    \rho_S(t) = \Tr_{EM}\rho(t),
\end{equation}
where the trace is taken over the degrees of freedom of the electric field.

One step we wish to highlight is where one integrates the Heisenberg's equation of motion \eqref{eq:Heis} from some $t_0$ to $t$ and then plug that result back into the same Heisenberg's equation of motion to get:  
\begin{multline}\label{eqn:finalVNE}
    \frac{d}{dt}\rho_S(t) = \Tr_{EM}\big\{-\frac{i}{\hbar}[H_\omega(t),\rho(t_0)] \\-\frac{1}{\hbar^2}\int_{t_0}^{t}dt' [H_\omega(t),[H_\omega(t'),\rho(t')]]\big\}.
\end{multline}
This step is necessary in order for the following two approximations to yield a non-trivial result. The first of the two approximation is the Born-Oppenheimer approximation that states that the system is in a separable state: $\rho(t) = \rho_{EM}(t)\otimes\rho_S(t)$. The second is the Markovian approximation that states that the time dependence of the density operator can only depend on the current state of the system and not it's past: $\rho(t'),\rho(t_0) \rightarrow \rho(t)$. 

The reason for this highlight is to emphasise the origin of the physical processes that happen when a two-level system is in contact with the EM (laser) field. The first term, linear in $H_\omega$, is responsible for Rabi oscillations characterised by the Rabi frequency $\Omega_{Rabi} = \frac{1}{\hbar}\boldsymbol{E}_0\cdot\boldsymbol{D}_{1,2} $, where $E_0$ is the expectation value of the amplitude of the oscillating electric (laser) field.
The second, quadratic, term is responsible for stimulated emission and spontaneous decay (incoherent damping of Rabi oscillations).


Performing the rest of the steps in deriving the Master equation and neglecting terms oscillating more rapidly than the detuning $\Delta = \omega_{12} - \omega$ (rotating-wave approximation), we arrive at the optical Bloch equations for the matrix elements of the reduced density matrix of the two-level system:

\begin{equation*}
    \frac{d\rho_{1,1}}{dt}= -i\frac{\Omega_{Rabi}}{2}\rho_{1,2} + i\frac{\Omega_{Rabi}^*}{2}\rho_{2,1} + \Gamma \rho_{2,2},
\end{equation*}

\begin{equation*}
    \frac{d\rho_{1,2}}{dt}= -i\frac{\Omega_{Rabi}^*}{2}(\rho_{1,1}-\rho_{2,2}) +(i\Delta- \frac{\Gamma}{2}) \rho_{1,2},
\end{equation*}

\begin{equation}
    \rho_{22} = 1- \rho_{11} \quad \text{and} \quad \rho_{12} = \rho_{21}^*,
    \label{eqn:opt}
\end{equation} 
where $\Gamma = \frac{1}{\tau}$ is the inverse lifetime of the excited state. The population of the excited state, $\rho_{2,2}$, can be probed by observing the spontaneous emission intensity in the reference experiments or, for a higher sensitivity, it could be measured through resonant ionization. \cite{gar18}. As $\comm{\hat{U}_\omega(t)}{\hat{U}_{\omega'}(t)} = 0$, these equations can be used in the case of time-dependent detuning by replacing $\Delta$ with $\Delta_D(t)$. The spatial dependence of the electromagnetic field can be introduced as a phase factor on the correlation coefficients. Thus, it does not contribute to the observed populations \cite{wannstrom1990optical}.
Ramsey protocols can be studied by using a time-dependent Rabi frequency in the equations above. 
\section{Doppler switching through different potential barriers} \label{sec:dopsw}

The population of the excited state of a two-level system interacting with a laser field near resonance was studied by using different potential barriers (see Figure \ref{fig:protocols}). The equations described in section \ref{sec:theory} were numerically integrated for different system configurations in order to study their performance in a Ramsey-like experiment. The value of $\Gamma$ was obtained from the inverse sum of Einstein's A coefficients for a relevant state, which are taken from the NIST database \cite{nist}.
Firstly, we test the model description with existing experimental results. A satisfactory description of previous observations allows us to study and propose alternative spectroscopy protocols.

\subsection{One potential barrier}
\begin{figure}[H]
  \includegraphics[width=\linewidth]{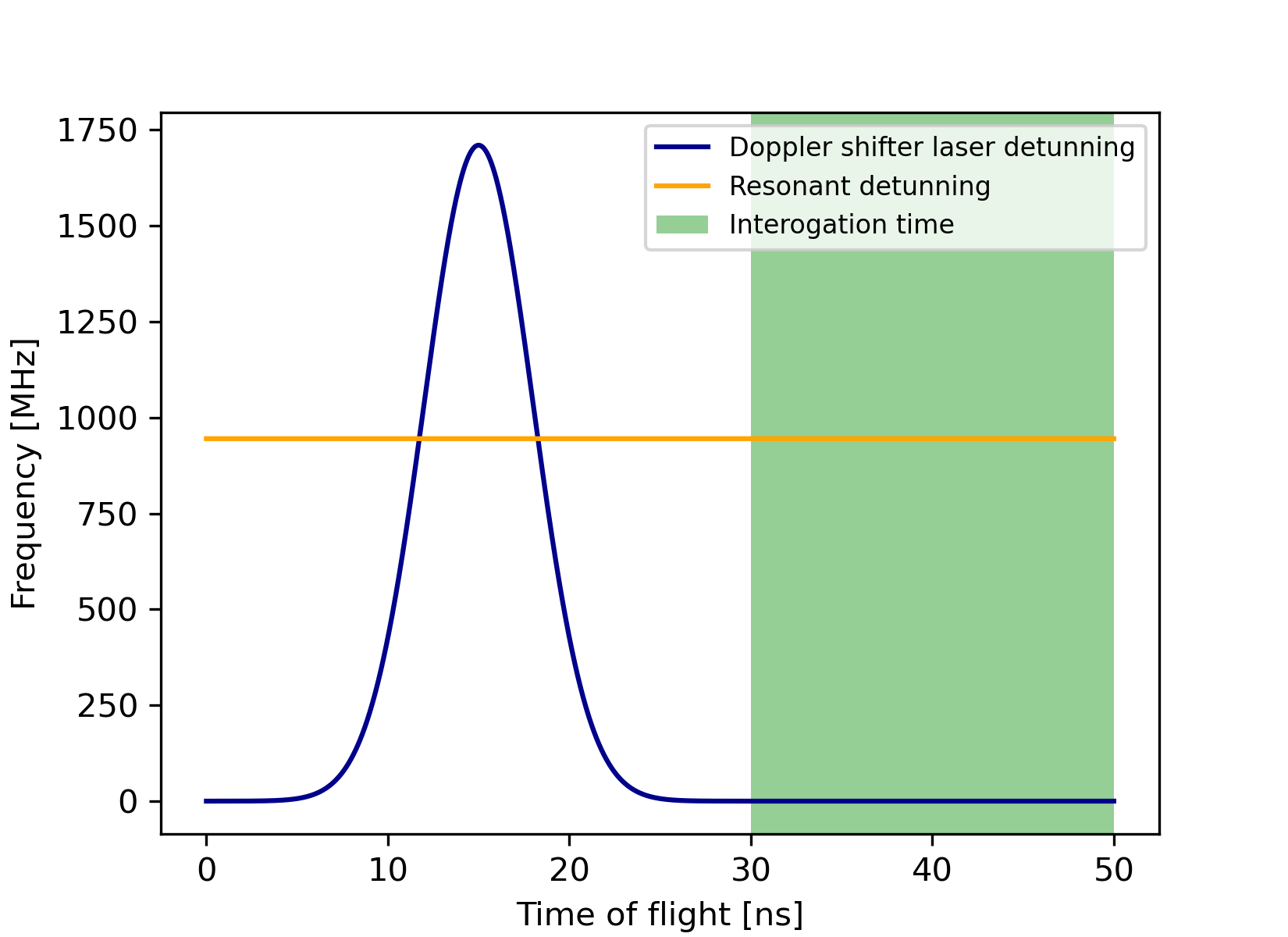}
  \caption{Scheme of Doppler switching with one potential barrier. The Doppler-shifted frequency as a function of time of flight is shown in blue \dgr. The resonant laser detuning needed to drive the transition is shown in orange \lgr. The frequency is shown relative to the laser frequency as seen in the ions' rest frame in the regions free from the electrostatic field. The green \dgr region shows interrogation time over which the excited population is integrated.}
  \label{fig:singleDprotocol}
\end{figure}

\begin{figure}[H]
\includegraphics[width=.9\linewidth]{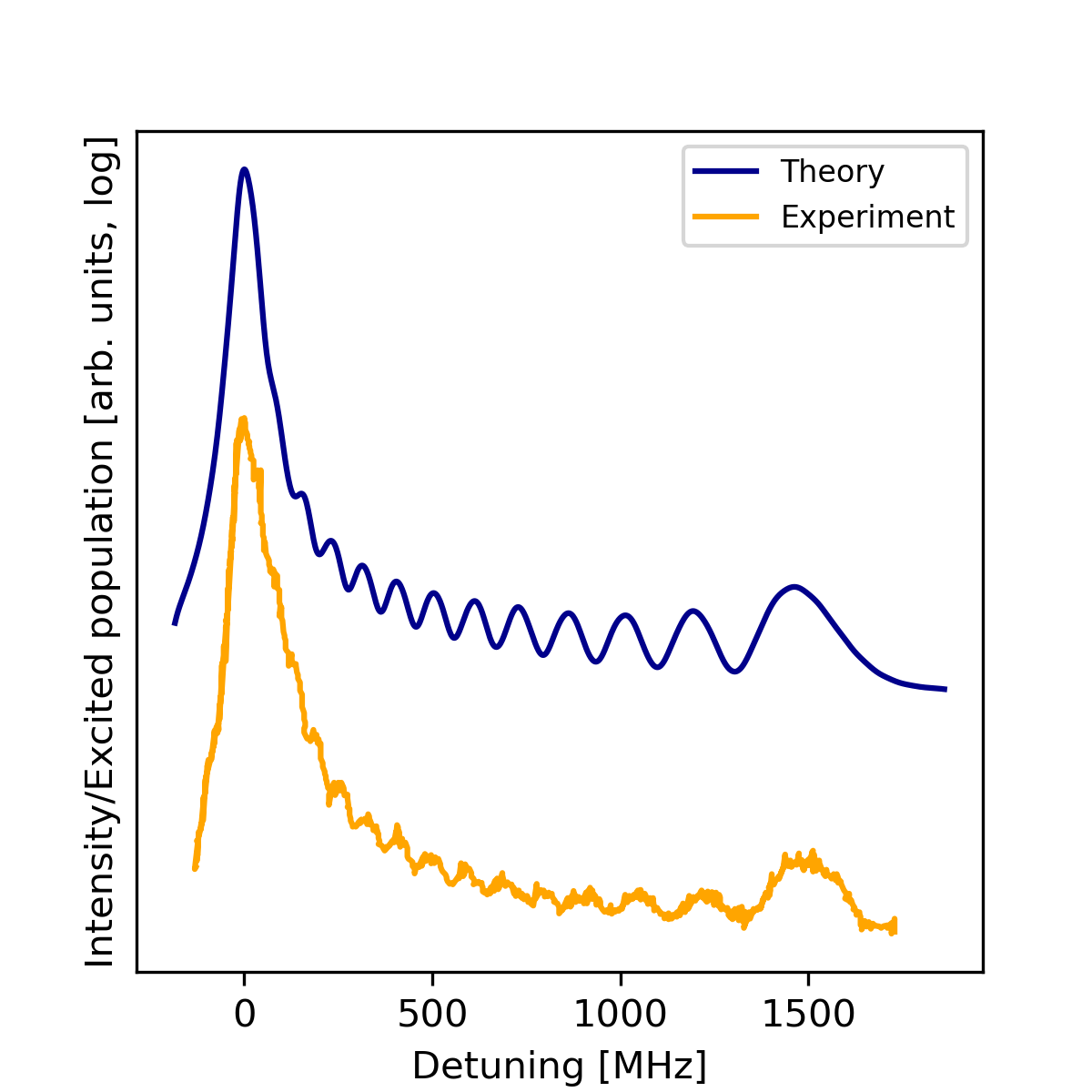}
\caption{Excited state population of an ion passing through one potential barrier.  The amplitude of the potential barrier was fixed, and the laser frequency was scanned around the Doppler-shifted resonance frequency. The experimental results from Ref. \cite{sil81} are compared with our theoretical results. The curves on the log-linear plot are offset on purpose, the overall multiplicative normalisation factor is setup dependent, hence, in this paper, arbitrary.}
\label{fig:exp1}
\end{figure}

 Ramsey fringes in collinear laser spectroscopy have been experimentally observed by using one potential barrier \cite{sil81,bor81}.  The experimental scheme is shown in Figure \ref{fig:singleDprotocol}. Using a fixed potential, the laser was scanned around the Doppler-shifted resonant frequency. The population of the excited state was probed by detecting fluorescent photons in a detector placed after the potential barrier. The experimental results from Ref. \cite{sil81,bor81} are compared with our numerical results in Figure \ref{fig:exp1}.
 
 The radiation intensity ($\sim E_0$) was obtained from the laser power and ion beam diameter reported in \cite{sil81}. The dipole matrix element, $e a_B$, was adjusted to give an order of magnitude estimate for Rabi frequency of 100 MHz (this is tunable in any setup). The geometry of the experimental schemes was taken as reported in Refs. \cite{sil81, bor81}. Laser detuning with respect to the probed transition was chosen to fit the experimental data and illustrate certain features thereof. The results are shown for the 455 nm transition line of Ba II ($6s ^2{S}_{1/2}$ $\rightarrow$ $6p ^2{P}_{3/2}$). The Einstein's A coefficient ($\sim 10^8 s^{-1}$ MHz) for this transition was taken from \cite{karlsson1999revised}. 
 The electric field produced by the potential barrier for this geometry was simulated using the software package COMSOL. The simulated electrostatic potential was used to provide a realistic estimation of the velocity distribution of the ions.

As seen in Figure \ref{fig:exp1}, a good agreement between theory and experiment is observed. Previous analysis neglected the lifetime of the excited state, the linewidth of the laser, and did not employ a realistic velocity distribution \cite{sil81,bor81}. In our current model, the lifetime was directly accounted for in the equation of motion and the laser linewidth was included through convolution with a Lorentzian lineshape. This approach is valid when the time of flight is comparable to the Rabi period, meaning that during that time the ion interacts with only one mode of the EM field. The linewidth of the laser is relevant when we have an ensemble of ions interacting with different modes of the laser light. Hence, we add contributions from different modes, i.e. by convolution of the frequency scan with the lineshape. Thermal effects were neglected. This scheme provides a test of the simple theoretical description, but is not suitable for Ramsey spectroscopy since scanning through interaction times by changing the velocity of the atoms would in-turn change the detuning.

\subsection{Two potential barriers}

\begin{figure}[H]
  \includegraphics[width=\linewidth]{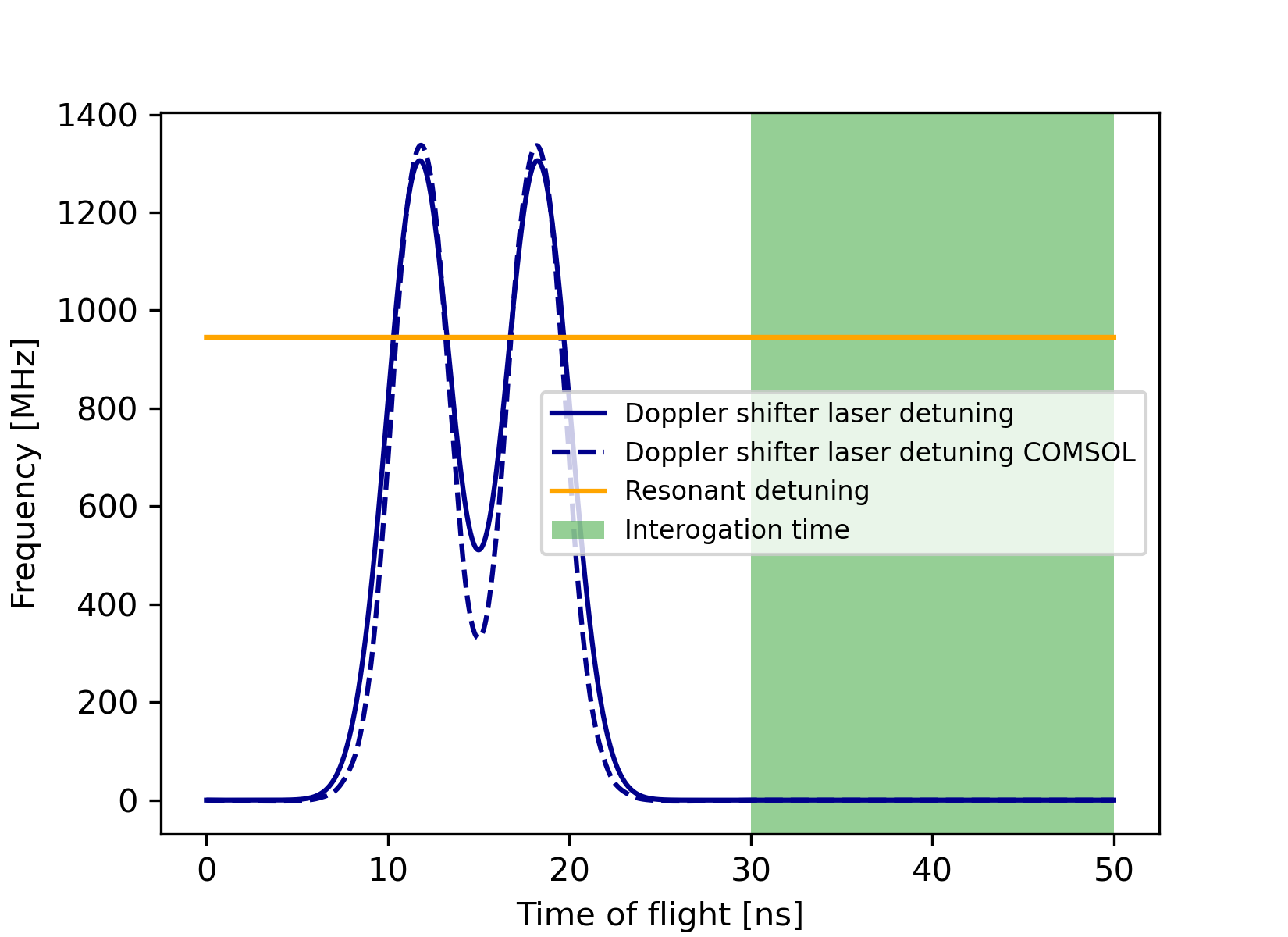}
  \caption{
     Doppler switching with two potential barriers: Dependence of Doppler-shifted frequency on the time of flight.  The resonant frequency as referenced from the Doppler-shifted laser frequency outside the electrostatic field is shown  (blue \dgr solid, blue \dgr dashed).    The green \dgr region shows interrogation time over which the excited population is integrated.
   The detuning over time (blue \dgr dashed) was modelled by simulating the electrostatic potential with the geometry and potentials of the setup reported in \cite{sil81}. A sum of two Gaussians (blue \dgr solid) was used as an approximation to simplify our simulations.
    }
  \label{fig:doubleDprotocol}
\end{figure}

In this section we study the population of the excited state of a two-level system when it is passed through two different potential barriers. The experimental scheme is shown in Figure \ref{fig:doubleDprotocol}. Experimental results for this configuration were presented in \cite{sil81}, but the observed lineshapes were not satisfactorily explained. The parameters of the model were obtained as explained in the previous section. Similarly, the electric potentials of the geometry stated in Ref. \cite{sil81} was simulated in COMSOL, and the simulated electric fields were used to provide a realistic distribution of the ion beam velocities.

\begin{figure}[H]
\centering
\includegraphics[width=.9\linewidth]{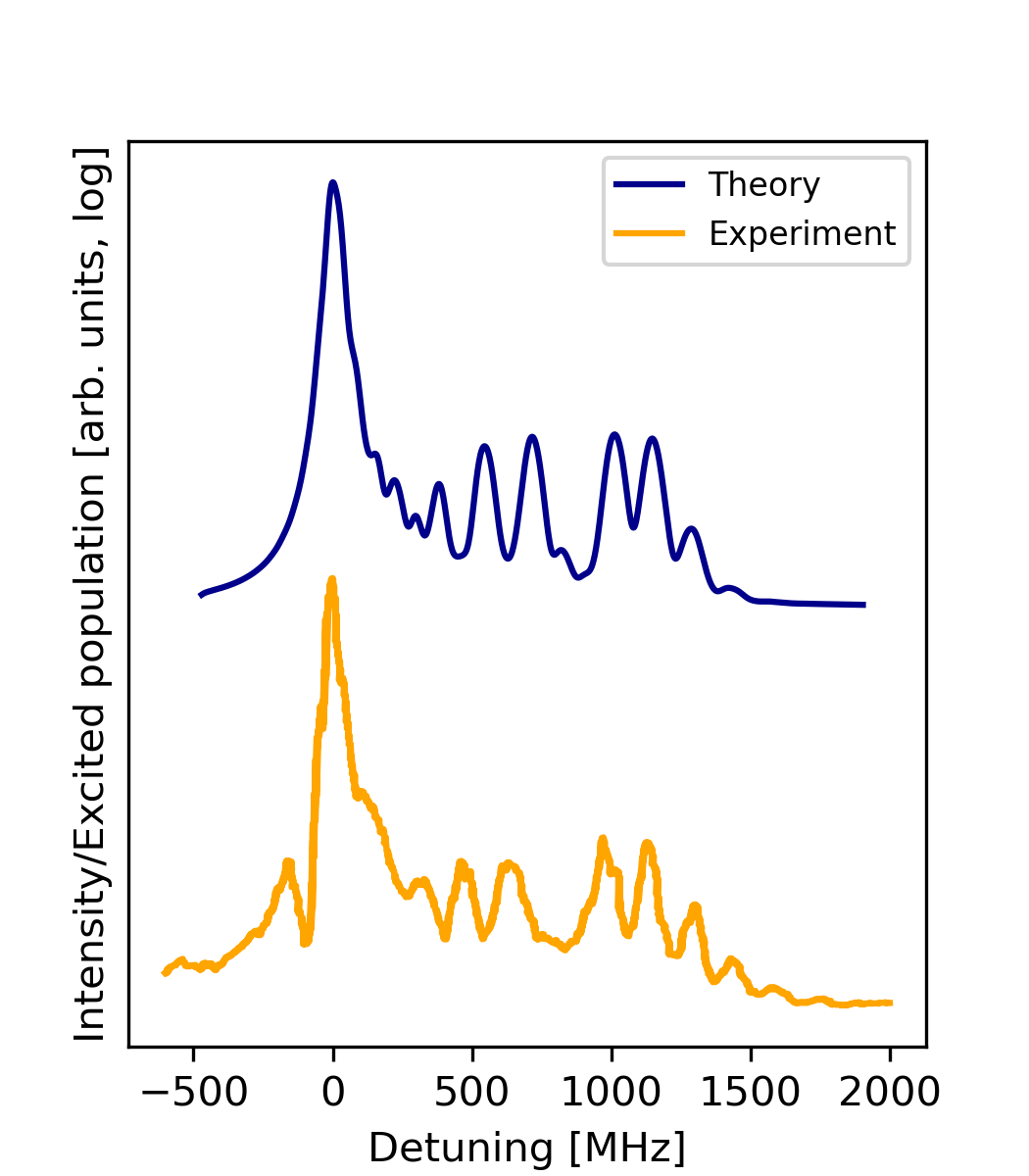}
\caption{Excited state population of an ion passing through two potential barriers. The amplitude of the potential barriers were fixed, and the laser frequency was scanned around the Doppler-shifted resonance frequency. The experimental results from Ref. \cite{sil81} are compared with our simulations. The curves on the log-linear plot are offset on purpose, the overall multiplicative normalisation factor is setup dependent, hence, in this paper, arbitrary.}
\label{fig:exp2}
\end{figure}

The experimental and theoretical results are presented in Figure \ref{fig:exp2}. A good agreement between the experimental lineshapes \cite{sil81} and our simulations is observed. Previous attempts to explain these lineshapes did not provide a complete description \cite{sil81}. We noticed that in addition to the model differences, the simulated electric fields were significantly different to the assumed functional forms used in Ref. \cite{sil81,bor81}.

In the experiment, the laser detuning was being scanned. This is illustrated in Figure \ref{fig:doubleDprotocol} where the position of the orange \lgr line, i.e. the value of the laser detuning relative to the resonance, can be varied. This is equivalent to changing the position along the ions' path where the laser frequency is on resonance with the ionic transition. These points correspond to the intersections of the blue \dgr and orange \lgr lines in Figure \ref{fig:doubleDprotocol}. On the same figure we see that depending on the detuning there can exist two or four positions at which the ion is on resonance with the laser field.
In the case where there are four points, the ions effectively see two potential barriers. This double switch case exhibits Ramsey oscillations at the fixed frequency corresponding to the separation of the two peaks.
When the detuning (orange \lgr curve) is lower than some critical value (around 800 MHz in this case),  only two intersections occur between the blue \dgr and orange \lgr lines. This case is similar to a single potential barrier discussed in the previous section. The ions do not "see" the dip of the frequency in the centre since it is detuned from resonance. These two different features are present in the experimental data \cite{sil81} shown in Figure \ref{fig:exp2}.

It is important to note that in both single and double barrier cases, the main Rabi-type peaks at no detuning in Figures \ref{fig:exp1} and \ref{fig:exp2}, are not broken up by Ramsey fringes. Hence, these protocols offer no improvement on the precision of determining the resonance position, as it is seen in the stationary Ramsey spectroscopy.

We will offer a protocol that can improve the precision in the following sections.

\section{Time Scanning with two potential barriers}
The good agreement of our simple model for one and two potential barriers, motivates us to explore the use of transient field phenomena to perform future precision studies. Here we propose a measurement protocol to implement Ramsey spectroscopy, using two Gaussian potential barriers separated by a variable distance. In this configuration, the population of the excited state can be measured as a function of the distance, and the frequency detuning can be obtained from the Fourier transform of the observed signal \cite{ramsey1950molecular}.

\begin{figure}[H]
  \includegraphics[width=\linewidth]{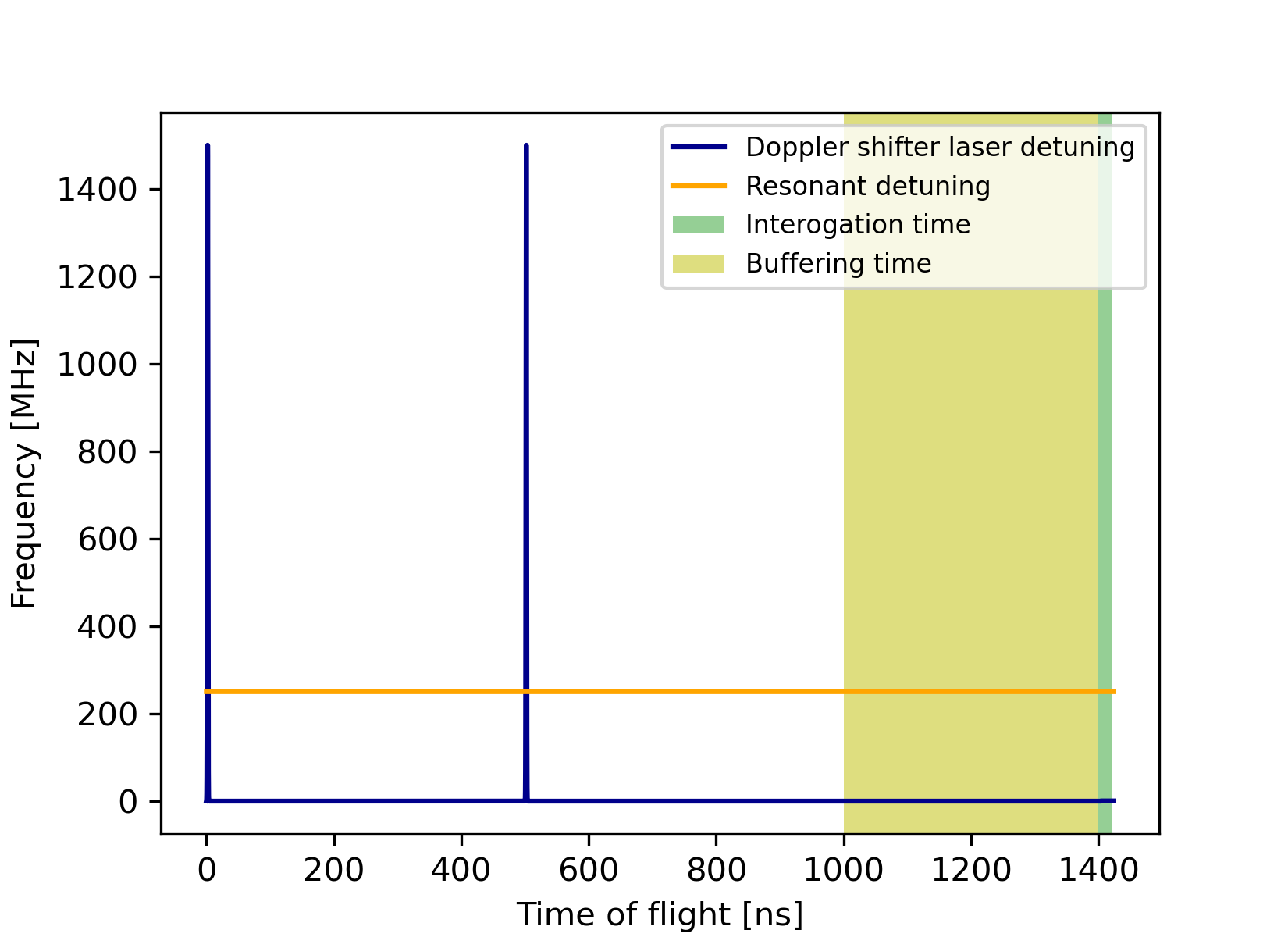}
  \caption{Doppler switching with two potential barriers of variable distance. The dependence of Doppler-shifted frequency as a function of time of flight is shown with the blue \dgr curves. The separation between two voltage pulses can be varied. The resonant frequency as referenced from the Doppler-shifted laser frequency outside the electrostatic field is marked with the orange \lgr curve. The green \dgr region shows the interrogation time over which the population of the excited state is integrated. }
  \label{fig:VarProt}
\end{figure}

The protocol for the proposed experimental scheme is shown in Figure \ref{fig:VarProt}. The distance between the two potential barriers (blue \dgr peaks) can be varied, and the population of the excited state can be measured as a function of the time-of-flight separation (Ramsey oscillations). This scheme can be implemented as a layered structure of concentric rings stacked on top of one another, similar to the setup presented in \cite{sil81} but at smaller scale. Such an electrode array can be manufactured by means of 3D printing \cite{jenzer2017study, jenzer2018study}. Assuming a period of the layered structure of about 0.5 mm over a 0.5 m span, a resolution of 1 kHz and bandwidth of the 1 GHz can be achieved for an ion mass of $A=100$ accelerated to 20 keV. 

\begin{figure}[H]
\begin{subfigure}{.5\textwidth}
  \centering
  \includegraphics[width=.8\linewidth]{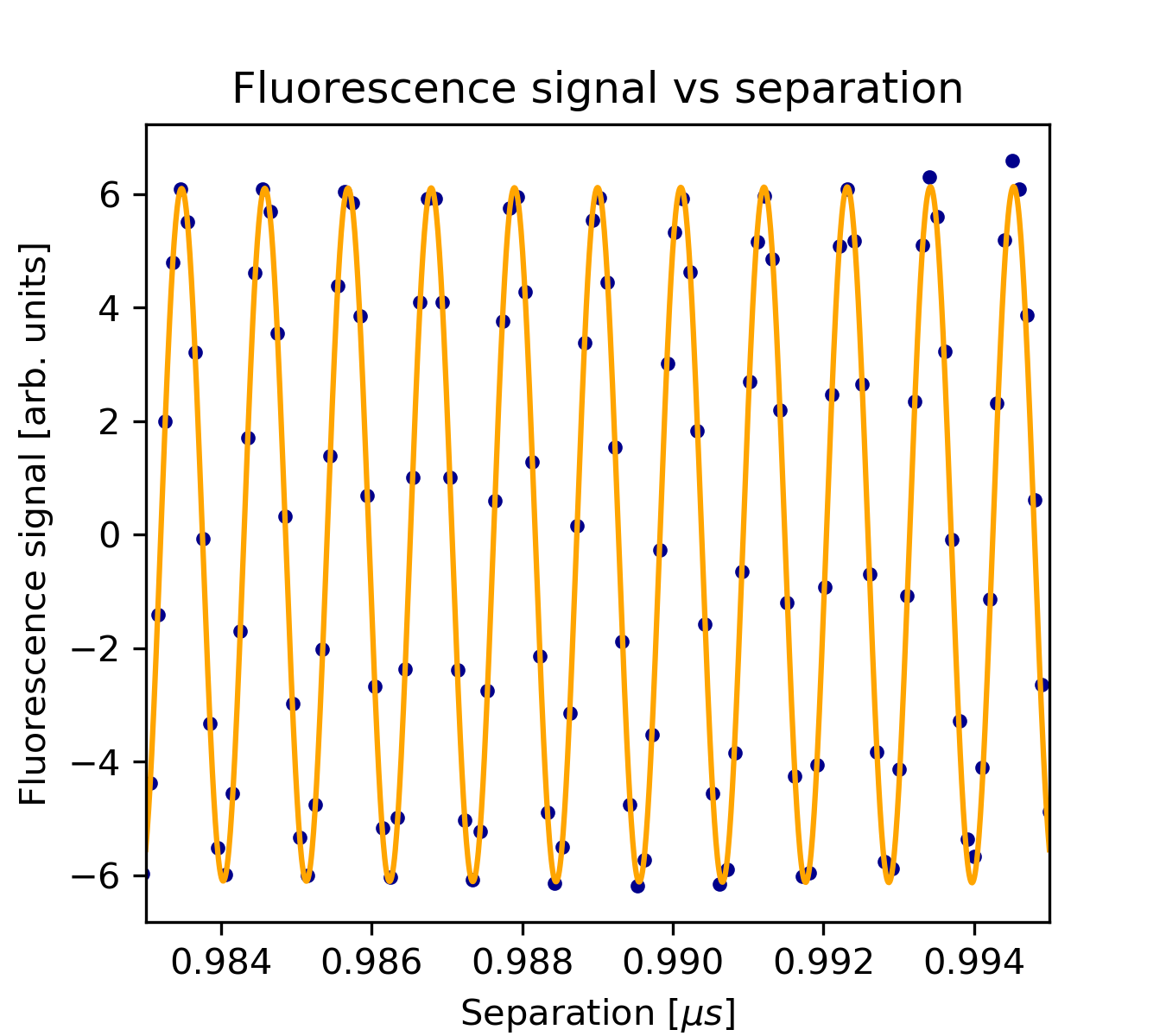}  
    \label{fig:sub-first}
\end{subfigure}
\begin{subfigure}{.5\textwidth}
  \centering
  \includegraphics[width=.8\linewidth]{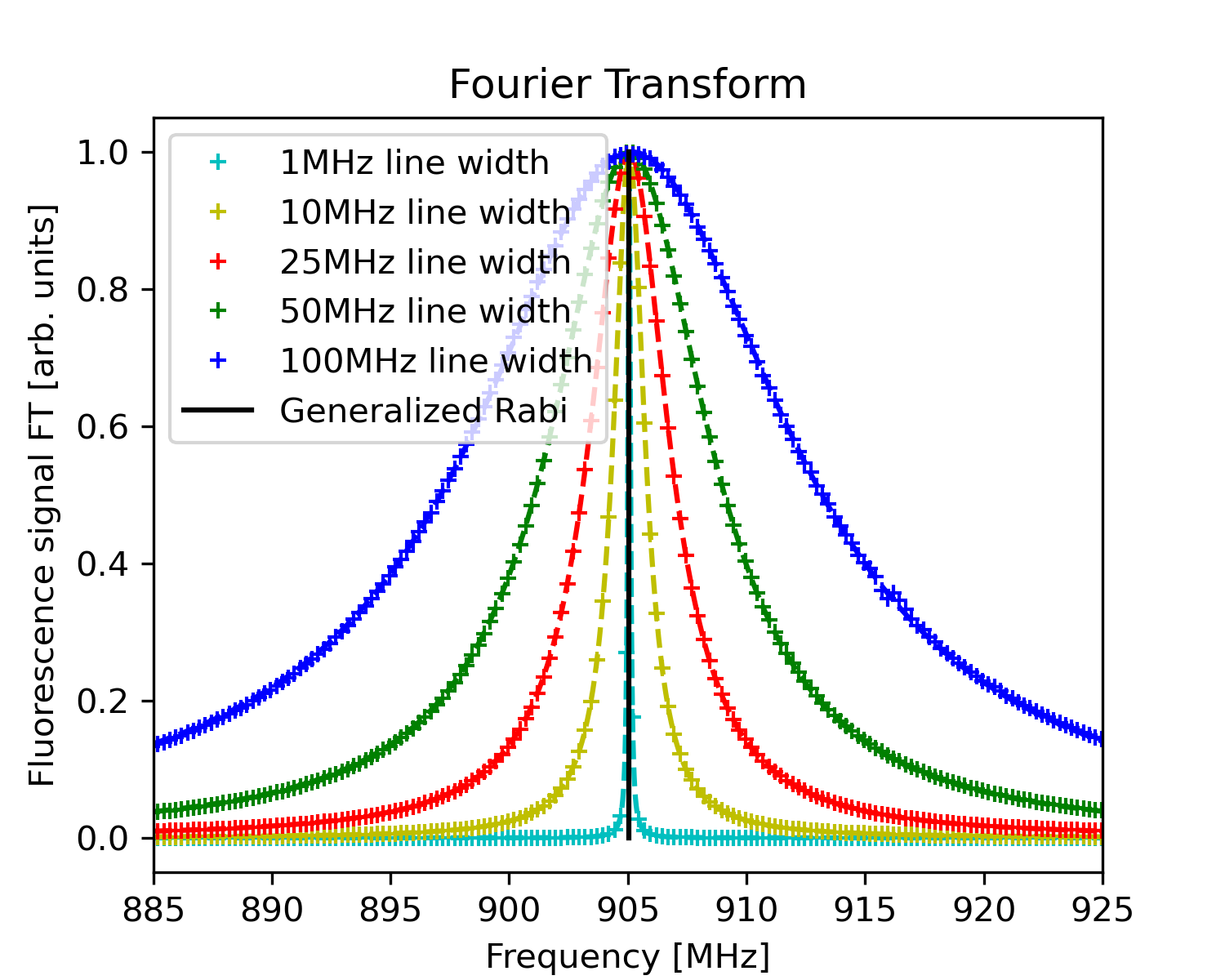}  
  \label{fig:sub-second}
\end{subfigure}
\begin{subfigure}{.5\textwidth}
  \centering
  \includegraphics[width=.8\linewidth]{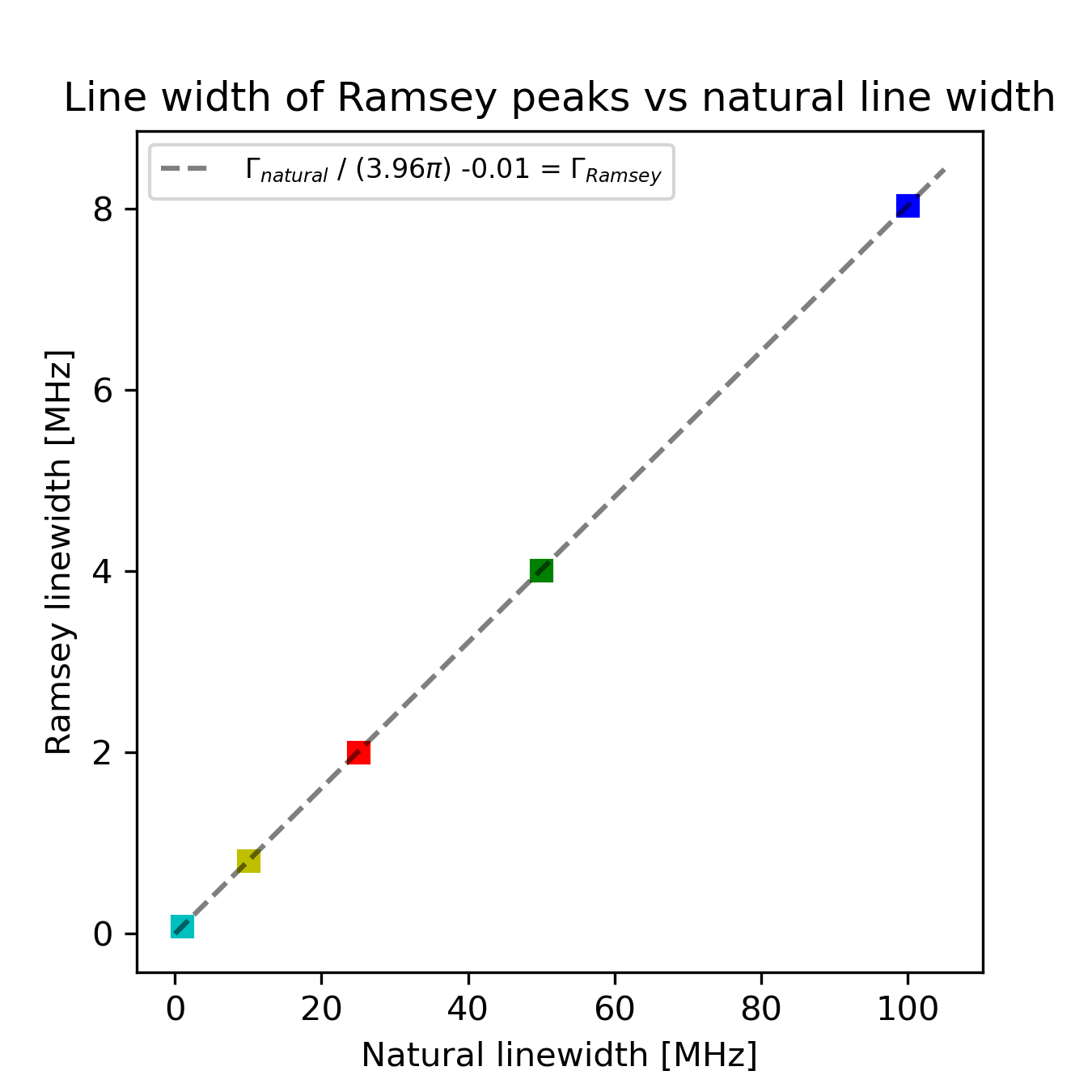}  
    \label{fig:sub-third}
\end{subfigure}

\caption{ Results from the time domain scan:   a) Ramsey oscillations seen in the fluorescent signal versus time between pulses, the points shown are taken from an arbitrary narrow window of the total 1000 ns simulated range. The oscillations seen are in agreement with the generalised Rabi frequency of our setup and the decay rate is half of the natural linewidth. The experimental points and the decaying exponential fit are shown in blue \dgr and orange \lgr respectively; b) Fourier transform of the Ramsey oscillations for different natural linewidths. The dashed line shows the Lorentzian fit; c) resolution obtained from the fit of the Ramsey fringes to Lorentzian shapes. The width of the Lorentizan shape, Ramsey linewidth, is shown as a function of the natural linewidth. }
\label{fig:t_scan}

\end{figure}

The simulations of the proposed scheme were performed using the following parameters: a pulse length, the time ions experience the electrostatic field of the barrier, of 0.1 ns; a pulse height, the maximal Doppler shift experienced by the atoms as they slow down while crossing the barrier, of 1880 MHz; a detuning of 900 MHz; a natural linewidth of 1 MHz; a Rabi frequency of 600 Mrad/s; a range of separation from 0 up to 1000 ns. The population of the excited state was obtained as an average over 200 ns, and assumed to be measured 250 ns after the last potential barrier.
The Ramsey oscillations simulated for the population of the excited state are shown in Figure \ref{fig:t_scan}a. The Fourier transformed of the signal is shown in Figure \ref{fig:t_scan}b. This procedure measures the generalised Rabi frequency, $\Omega_{GEN}$, which is related to the frequency detuning from resonance, $\Delta$, and the Rabi frequency $\Omega_{Rabi}$
\begin{equation}
\Omega_{GEN}^2 = \Delta^2+\Omega_{Rabi}^2. \label{gen_rabi}
\end{equation}
High laser powers can cause a detuning from the laser frequency due to the AC Stark-like shift, without introducing power broadening, which can be a major disadvantage in Rabi-type measurement. 

We have not modelled noise in this study and the parameters we chose were arbitrary, however, the following two facts are relevant for the actual experimental setup: i. the area under the pulse in the detuning profile measured in the time-of-flight domain determines the visibility of the fringes (the AC amplitude divided by the DC level of the fluorescence); ii. the Rabi frequency determines the magnitude of the overall signal (both DC and AC). Hence the actual signal-to-noise ratio heavily depends on the details of a particular experiment.

The lineshape of the Ramsey oscillations was studied using the same model parameters and scanning from 200 ns to 4200 ns. The simulations were performed assuming different natural linewidths. The result of this analysis is represented in Figure \ref{fig:t_scan}. A Lorentzian curve fits the lineshape well, with a centroid that is within 10kHz (for the narrowest 1kHz) of the expected value. Fitting the time domain can offer sub-kHz agreement across all linewidths.  As can be seen from equation (\ref{eqn:opt}), Ramsey lines are twice as narrow than the equivalent lineshapes obtained for Rabi-type measurements. The coherences', which are responsible for the oscillations, decay time is twice as long as populations' decay time.

\begin{figure}[H]
	\includegraphics[width=\linewidth]{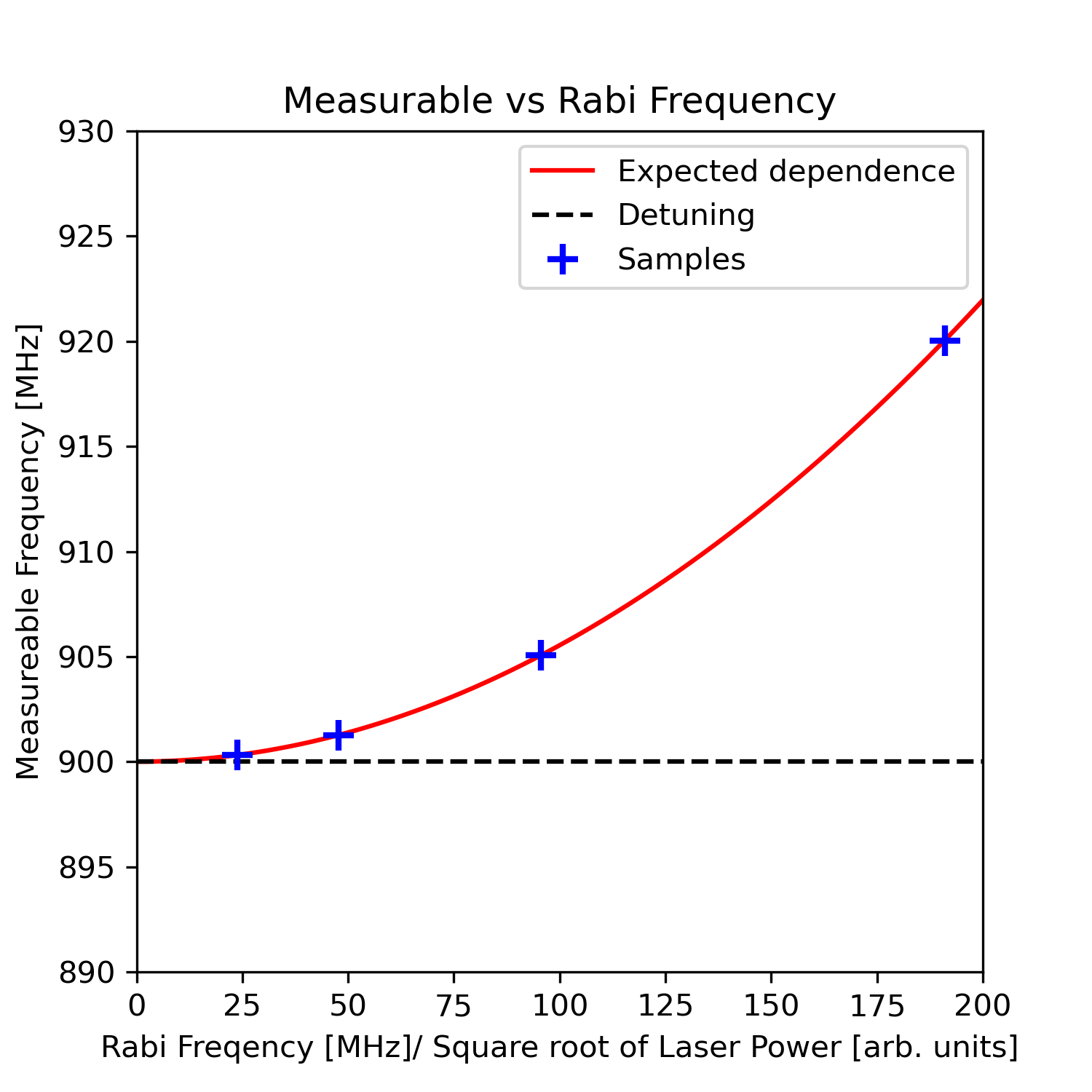}
	\caption{Relationship between the Ramsey frequency and Rabi frequency. The solid line representing the expected dependence. }
	\label{fig:AC Stark}
\end{figure}

The AC Stark-like shift can be corrected by measuring the frequency at different laser powers. This procedure is illustrated in Figure \ref{fig:AC Stark}. Frequency measurement with different powers of the probing laser, $P$, can be expressed as $P= \chi^{-1} \Omega_{Rabi}^2$, with $\chi$ an experimentally dependent proportionality constant that can be obtained by fitting the function $f(P) = \sqrt{\Delta^2 + \chi P}$ to the experimental results.

\section{Summary and outlook} \label{sec:conclusion}

Motivated by the potential use of transient phenomena to improve the spectral resolution in collinear laser spectroscopy, we present a simple description of a co-propagating laser interacting with an ion beam travelling through different potential barriers. This  model provides a good description of previous experimental results that were not fully understood. A experimental scheme was proposed with the aim of implementing
Ramsey-type measurements with fast ion beams for future precision studies. The proposed measurement protocol allows for varying the time-of-flight separation between potential barriers.

Two main advantages of the proposed transient interference protocol over the usual Rabi-type collinear spectroscopy can be highlighted: i. the detuning is measured by observing the coherences of the density matrix, which results in half of the linewidth; ii. power broadening is eliminated in favour of an AC Stark effect, which can be empirically obtained.

Our results neglected the thermal motion of the ions along the beam direction. Including this effect would be important for an accurate description of precision experiments. 

\begin{acknowledgments}
This work is supported by Department of Energy, Office of Science, Office of Nuclear Physics, under Award numbers DE-SC0021176 and DE-SC0021179. JJ is partly supported by the Oxford-ShanghaiTech collaboration agreement. We thank to Silviu Udrescu, Adam Vernon, and Shane Wilkins for fruitful comments and suggestions.  
\end{acknowledgments}



\FloatBarrier
\bibliography{main}

\end{document}